\documentclass[%
aip,  jmp, amsmath,amssymb,
reprint,%
]{revtex4-1}

\usepackage{graphicx}
\usepackage{dcolumn}
\usepackage{bm}

\begin{document}


\title{Coupled multiphysics, barrier localization, and critical radius effects in embedded nanowire superlattices} 



\author{Sanjay Prabhakar}
\email[]{sprabhakar@wlu.ca}
\homepage[]{http://www.m2netlab.wlu.ca}
\affiliation{M\,$^2$NeT Laboratory, Wilfrid Laurier University, 75 University Avenue West, Waterloo, ON, Canada, N2L 3C5
}
\author{Roderick Melnik}
\affiliation{M\,$^2$NeT Laboratory, Wilfrid Laurier University, 75 University Avenue West, Waterloo, ON, Canada, N2L 3C5
}
\affiliation{
Gregorio Millan Institute, Universidad Carlos III de Madrid, 28911, Leganes, Spain
}
\author{Luis L Bonilla}
\affiliation{
Gregorio Millan Institute, Universidad Carlos III de Madrid, 28911, Leganes, Spain
}

\date{April 30, 2013}

\begin{abstract}
The new contribution of this paper is to develop a cylindrical representation of an already known multiphysics model
for  embedded nanowire superlattices (NWSLs) of wurtzite structure that includes a coupled, strain dependent  8-band $\mathbf{k\cdot p}$ Hamiltonian in cylindrical coordinates and investigate the influence of coupled piezo-electromechanical effects on the barrier localization and critical radius in such NWSLs. The coupled piezo-electromechanical model for semiconductor materials takes into account the strain, piezoelectric effects and spontaneous polarization. Based on the developed 3D model, the band structures of electrons (holes) obtained from results of modeling in Cartesian coordinates are in good agreement with those values obtained from our earlier developed 2D model in cylindrical coordinates. Several parameters such as lattice mismatch, piezo-electric fields, valence  and conduction band offsets at the heterojunction of  $\mathrm{Al_xGa_{1-x}N/GaN}$ supperlattice can be varied as a function of the Al mole fraction. When the band offsets  at the heterojunction of $\mathrm{Al_xGa_{1-x}N/GaN}$ are very small and the influence of the piezo-electromechanical effects can be  minimized, then the barrier material can  no longer be treated as an infinite potential well. In this situation, it is possible to visualize  the penetration of the Bloch wave function  into the barrier material that provides an estimation of  critical radii of NWSLs. In this case,  the NWSLs can act as inversion layers. Finally, we investigate the influence of symmetry of the square and cylindrical NWSLs on the band structures of electrons in the conduction band.  However for larger lateral size of the NWSLs, the influence of such edge effects on the band structures of the NWSLs  are not substantially influenced by the choice of either square or cylindrical
symmetry but the localization of  weavefunction with square symmetry  is different than for the case with cylindrical symmetry  which might indicate that the symmetry is broken in square shape NWSLs.
\end{abstract}


\maketitle 

\section{Introduction}
Studies on low dimensional systems, such as nanowires and superlattices, have attracted considerable attention, spurred on by the development of smaller and faster electronic devices and by the exploitation of their extraordinary properties for improved performance in various areas of science and technology, including nano- and micro-electronics, thermoelectricity and magnetism.~\cite{huang01,lin03,venkatasubramanian01,harman02,nielsch01,prabhakar13} Today's technology allows finite length modulated quantum wire heterostructures to be grown in what is known as NWSLs. NWSLs are the nanoscale building blocks that through the bottom up assembly can enable diverse applications. One can expect a straightforward analogy to the planar electronic/optoelectronic industry to extrapolate that complex compositionally  modulated superlattice structures could greatly increase the versatility and power of these building blocks in nanoscale applications.~\cite{huang02}

In the NWSLs, the localization in  barriers and the critical radius are important issues.~\cite{chuang07,cirlin09,raychaudhuri06} The existence  of barrier localization and the calculation of critical radius in NWSLs have been previously carried out by using one band effective mass theory in Refs.~\onlinecite{voon04,willatzen04}. Those earlier  results indicated the possibility of these modulated structures to display free carrier like behavior along the nanowire axis when a critical wire radius is considered.  Moreover, it has been shown that the existence of critical radii for inversion of state localization is a much more general phenomenon.  Most of these studies deal with free standing NWSLs. The barrier localization and critical radius phenomena can be particularly important in AlGaN/GaN NWSLs where band structure parameters can be controlled with the variation of Al mole fraction.
In this situation, the influence of piezo-electromechanical effects can be minimized. As a result, we see the penetration of the wavefunctions into the barrier materials which provides the evidence of the presence of critical radii in the NWSLs.~\cite{voon04,voon03}

\section{Piezo-electromechanical effects}

\subsection{Piezo-electromechanical effects in Cartesian coordinates}
To investigate the influence of piezo-electromechanical effect on the band structure calculation of low dimensional semiconductor nanostructures, following Refs.~\onlinecite{lassen08}, first we write the coupled system of the Navier equations for stress and Maxwel's equations for piezoelectric fields as~\cite{melnik00,prabhakar12b}
\begin{eqnarray}
\partial_j \sigma_{ik}=0,   \label{del-j}\\
\partial_i D_i=0.  \label{del-i}
\end{eqnarray}
The stress tensor components $\sigma_{ik}$ and the electric displacement vector components $D_i$  can be written as~\cite{barettin08,lassen08}
\begin{eqnarray}
\sigma_{ik}=C_{iklm}\varepsilon_{lm}+e_{nik}\partial_nV,  \label{sigma-ik}\\
D_i=e_{ilm}\varepsilon_{lm}-\hat{\epsilon}_{in} \partial_n V + P_{sp}\delta_{iz}, \label{D-i}
\end{eqnarray}
where $C_{iklm}$ are the elastic moduli constants,   $e_{ik}$ is the piezoelectric constant, $\epsilon_{in}$ is the permittivity, $V$ is the piezoelectric potential, $P_{sp}$ is the spontaneous polarization  and $V$ is the built in piezoelectric potential. Also, $\varepsilon_{ik}$  are the components of strain tensors which are written as
\begin{equation}
\varepsilon_{ij}=\varepsilon_{ij}^u+\varepsilon_{ij}^0, \label{varepsilon-ij}
\end{equation}
where $\varepsilon_{ij}^0$ are the local intrinsic strain tensor components due to lattice mismatch and $\varepsilon_{ij}^u$ is position dependent strain tensor components. These two can be written as
\begin{eqnarray}
\varepsilon_{ij}^0=\left(\delta_{ij}-\delta_{iz}\delta_{jz}\right)a + \delta_{iz}\delta_{jz}c,\label{varepsilon-ij-0}\\
\varepsilon_{ij}^u=\frac{1}{2}\left(\partial_ju_i+\partial_iu_j\right),\label{varepsilon-ij-u}
\end{eqnarray}
where $a=\left(a_0-a\right)/a_0$ and  $c=\left(c_0-c\right)/c_0$ are the local intrinsic strains along a- and c-directions, respectively
(which are nonzero in the quantum well and zero otherwise).  Here, $a_0$, $c_0$ and  $a$,  $c$ are the lattice constants of the quantum well  and the barrier material of the NWSLs.

\subsection{Piezo-electromechanical effects in cylindrical coordinates}
In cylindrical polar coordinates ($r,\phi,z$), Eqs.~(\ref{sigma-ik}) and~(\ref{D-i}) can be written as~\cite{melnik00,barettin08,prabhakar10,patil09}
\begin{eqnarray}
\sigma_{rr}=C_{11}\varepsilon_{rr} + C_{12}\varepsilon_{\phi\phi} + C_{13}\varepsilon_{zz}+e_{31}\partial_z V,\label{sigma-rr-2}\\
\sigma_{\phi\phi}=C_{11}\varepsilon_{\phi\phi} + C_{12}\varepsilon_{\rho\rho} + C_{13}\varepsilon_{zz}+e_{31}\partial_z V,\\
\sigma_{r z}=2 C_{44}\varepsilon_{r z} + e_{15}\partial_r V,\\
\sigma_{zz}=C_{13}\varepsilon_{rr} + C_{13}\varepsilon_{\phi\phi} + C_{33}\varepsilon_{zz}+e_{33}\partial_z V,\\
D_r=2 e_{15}\varepsilon_{r z}-\epsilon_1 \partial_r V,\\
D_z= e_{31}\left(\varepsilon_{rr}+\varepsilon_{\phi\phi}\right)e_{33}\varepsilon_{zz}-\epsilon_3 \partial_z V + P_z^{sp}.
\label{sigma-rr-1}
\end{eqnarray}

Also, the coupled equations of wurtzite structure in the presence of piezo-electromechanical effects  in cylindrical coordinates can be written as~\cite{melnik00}
\begin{eqnarray}
\partial_r \sigma_{rr}+\partial_z \sigma_{r z}+\frac{\sigma_{rr}-\sigma_{\phi\phi}}{r}=0,\label{sigma-rra}\\
\partial_r \sigma_{r z}+\partial_z \sigma_{zz}+\frac{1}{r}\sigma_{r z}=0,\label{sigma-rrb}\\
\partial_r D_r+\partial_z D_z+\frac{1}{r} D_r=0.
\label{sigma-rr}
\end{eqnarray}
From Eqs.~\ref{sigma-rr-2} to~\ref{sigma-rr-1}, the components of the strain tensor, expressed through the displacement vector $\mathbf{u}=\left(u_r,u_\phi,u_z\right)$, for which $\partial_\phi \mathbf{u}=0$, are
\begin{eqnarray}
\varepsilon_{rr}=\partial_r u_r+ a, ~~~\varepsilon_{zz}=\partial_z u_z+c \label{varepsilon_rr}\\ \varepsilon_{\phi\phi}=\frac{ u_r}{r}+a,\, \varepsilon_{r z}=\frac{1}{2}\left(\partial_z u_r+\partial_r u_z\right).\label{varepsilon_rz}
\end{eqnarray}
The strain tensor components and the piezoelectric field (potential) can be found by solving the electroelasticity  problem~(\ref{sigma-rra}), (\ref{sigma-rrb}) and~(\ref{sigma-rr}).

\section{Band structure calculations}

\subsection{8-band $\mathbf{k\cdot p}$ model in Cartesian coordinates}
The steady state Schr\"odinger equation of the Kane model for the electrons in the conduction band and holes in the valence band  can be written as~\cite{kane57,bir74,lowdin51,rinke08,fu08,prabhakar12b}
\begin{equation}
\mathbf{H}\mbox{\boldmath$\psi$}=E\mbox{\boldmath$\psi$},
\label{H-prim}
\end{equation}
where
\begin{equation}
\mathbf{H}  =
\left(\begin{array}{cc}
\mathbf{H_c} & \mathbf{H_{cv}}  \\
\mathbf{H^{\dag}_{cv}} & \mathbf{H_v}  \\
\end{array}\right),
\mbox{\boldmath$\psi$}= \left(\begin{array}{c}
\mbox{\boldmath$\psi_c$}  \\
\mbox{\boldmath$\psi_v$} \\
\end{array}\right),
\label{H-prim-1}
\end{equation}
with $\mbox{\boldmath$\psi_c$}  = \mbox{\boldmath$\psi_c$}\left(\mathbf{r}\right)$ and $\mbox{\boldmath$\psi_v$} = \mbox{\boldmath$\psi_v$} \left( \mathbf{r}\right)$  are the position dependent conduction and valence band envelope functions.

The total wave function $\mathbf{\Psi}$ is:~\cite{winkelnkemper06,prabhakar12b}
\begin{equation}
\mathbf{\Psi} = \sum_{j=c,x,y,z} f_j \psi_j = \mbox{\boldmath$f$}\mbox{\boldmath$\psi$},
\label{Psi}
\end{equation}
where $\mbox{\boldmath$f$} = \left(f_c~ f_x~ f_y~ f_z\right)$ and $\mbox{\boldmath$\psi$}= \mbox{\boldmath$\left(\psi_c~ \psi_x~ \psi_y~ \psi_z \right)^{T}$}$.  The functions $\mbox{\boldmath$f$}$ are spinless and $\mbox{\boldmath$\psi$}$  is a spinor:
\begin{equation}
\mbox{\boldmath$\psi_j$} =
\left(\begin{array}{c}
\psi^1_j  \\
\psi^2_j\\
\end{array}\right), \quad \emph{j}=c,x,y,z.
\end{equation}
Hence, the basis functions of the Hamiltonian~(\ref{H-prim-1}) take the following form:~\cite{winkelnkemper06,prabhakar12b}
\begin{equation}
\left(f_c\psi_c^1,~f_c\psi_c^2,~f_x\psi_x^1,~f_x\psi_x^2,~f_y\psi_y^1,~f_y\psi_y^2,~f_z\psi_z^1,~f_z\psi_z^2\right)^T\nonumber
\end{equation}

We now turn to the description of the matrix Hamiltonian $\mathbf{H}$ of~(\ref{H-prim-1}). The diagonal element of the conduction band Hamiltonian $\mathbf{H_c}$ can be written as
\begin{eqnarray}\label{Hc}
H_c = &A'_1 k_z^2 + A'_2\left( k_x^2 +k_y^2 \right)+ U_c\nonumber\\
&+a_1 \varepsilon_{zz} + a_2 \left( \varepsilon_{xx} + \varepsilon_{yy}\right),
\end{eqnarray}
where $U_c = U_c\left(\mathbf r\right)$ is the position dependent edge of the conduction band $\Gamma_1$,  $a_1$ and $a_2$ are deformation potentials for the conduction band. The parameters $A'_1$ and $A'_2$ are expressed via the components $1/m_{\parallel}$ and $1/m_{\perp}$ of the tensor of the reciprocal effective masses for the conduction band in the single-band approximation and the Kane parameters $P_1 = -i \hbar \langle \phi_c \vert \hbar k_z \vert \phi_z \rangle/m_0$ and $P_2 = -i \hbar \langle \phi_c \vert \hbar k_x \vert \phi_x \rangle /m_0$. They are given by~\cite{winkelnkemper06,prabhakar12b}
\begin{eqnarray}
A'_1=\frac{\hbar^2}{2m_{\parallel}}-\frac{P^2_1}{E_g},\\
A'_2=\frac{\hbar^2}{2m_{\perp}}-\frac{P^2_2}{E_g},
\label{A-prim}
\end{eqnarray}
where $E_g$ is the band gap of semiconductor materials.

The intra-valence-band Hamiltonian $\mathbf{H_v}$ can be written as
\begin{equation}
\mathbf{H_v} = \mathbf{H^{(0)}} + \mathbf{H^{(so)}} +  \mathbf{H^{(\varepsilon)}} +\mathbf{ H^{\prime (k)}}.
\label{Hv}
\end{equation}
The Hamiltonian $\mathbf{H^{(0)}}$ entering Eq.~(\ref{Hv}) represents the position-dependent potential energy of an electron:
\begin{equation}
\mathbf{H^{(0)}} =
\left(\begin{array}{ccc}
U_{v6}&0&0\\
0&U_{v6}&0\\
0&0&U_{v1}
\end{array}\right),
\label{H0}
\end{equation}
where $U_{v6} = U_{v6}\left(\mathbf r\right)$ and $U_{v1} = U_{v1}\left(\mathbf r\right)$ are the position dependent edges of the valence bands $\Gamma_6$ and $\Gamma_1$, respectively.

The spin-orbit Hamiltonian $\mathbf{H^{(so)}}$ in  Eq.~(\ref{Hv}) can be treated as a perturbation term and can be written as~\cite{chuang96,bir74}
\begin{equation}
\mathbf{H^{(so)}} =i
\left(\begin{array}{ccc}
0&-\Delta_2 \sigma_z&\Delta_3 \sigma_y\\
\Delta_2 \sigma_z&0&-\Delta_3 \sigma_x\\
-\Delta_3 \sigma_y&\Delta_3 \sigma_x&0
\end{array}\right),
\label{Hso}
\end{equation}
where   $\Delta_2 = \Delta_2\left(\mathbf{r}\right)$ and $\Delta_3= \Delta_3\left(\mathbf{r}\right)$ are the parameters of the valence-band spin-orbit splitting and $\sigma_i (i=x,y,z)$ are the Pauli spin matrices:
\begin{equation}
\sigma_x=
\left(\begin{array}{cc}
0&1\\
1&0
\end{array}\right),~  \sigma_y=
\left(\begin{array}{cc}
0&-i\\
i&0
\end{array}\right),~  \sigma_z=
\left(\begin{array}{cc}
1&0\\
0&-1
\end{array}\right).
\label{sigma-xyz}
\end{equation}
The kinetic energy Hamiltonian $\mathbf{H^{\prime (k)}}$  in Eq.~(\ref{Hv}) can be written as~\cite{winkelnkemper06}
\begin{widetext}
\begin{equation}
\mathbf{H^{\prime(k)}} =
\left(\begin{array}{ccc}
L'_1 k^2_x + M_1 k^2_y + M_2 k^2_z & N'_1 k_x k_y & N'_2 k_x k_z \\
N'_1 k_x k_y & M_1 k^2_x + L'_1 k^2_y + M_2 k^2_z & N'_2 k_y k_z \\
N'_2 k_x k_z & N'_2 k_y k_z & M_3\left( k^2_x + k^2_y \right) + L'_2 k^2_z
\end{array}\right),
\label{Hprimek}
\end{equation}
\end{widetext}
where
\begin{eqnarray}\label{LNprimes}
L'_1 = L_1 + \frac{P^2_1}{E_g},~~~
L'_2 = L_2 + \frac{P^2_2}{E_g },\\
N'_1 = N_1 + \frac{P^2_1}{E_g },~~~
N'_2 = N_2 + \frac{P_1 P_2 }{E_g}.
\end{eqnarray}
Also,
\begin{equation}
\begin{split} \label{LMN_thru_A}
&L_1 = \frac {\hbar^2}{2m_0}\left( A_2 +A_4 +A_5 \right),\\
&M_1 = \frac {\hbar^2}{2m_0}\left( A_2 +A_4 -A_5\right),\\
&N_1 = \frac {\hbar^2}{2m_0} 2 A_5,
\quad L_2 = \frac {\hbar^2}{2m_0}A_1, \\
&M_2 = \frac {\hbar^2}{2m_0}\left( A_1+A_3 \right),
\quad N_2 = \frac {\hbar^2}{2m_0} \sqrt 2 A_6\\
&M_3 = \frac {\hbar^2}{2m_0} A_2, \quad N_3 = i \sqrt 2 A_7,
\end{split}
\end{equation}
with $A_1$, $A_2$, \ldots $A_7$ being real material parameters in conventional notations ~\cite{chuang96,bir74,winkelnkemper06}, $m_0$ is the free electron mass. Wurtzite structure has six fold rotational symmetry and thus we use the  relation $L'_1 - M_1 = N'_1$.
~\\
~\\
~\\
~\\
~\\
~\\

Finally, the strain tensor components are written as~\cite{fonoberov03}
\begin{widetext}
\begin{equation}\label{He}
\mathbf H^{(\varepsilon)} = \begin{pmatrix}
l_1 \varepsilon_{xx} + m_1 \varepsilon_{yy} + m_2 \varepsilon_{zz} &
n_1 \varepsilon_{xy} & n_2 \varepsilon_{xz}\cr
n_1 \varepsilon_{xy} &
m_1 \varepsilon_{xx} + l_1 \varepsilon_{yy} + m_2 \varepsilon_{zz} &
n_2 \varepsilon_{yz}\cr
n_2 \varepsilon_{xz} & n_2 \varepsilon_{yz} &
m_3\left( \varepsilon_{xx} + \varepsilon_{yy}\right) + l_2 \varepsilon_{zz}
\end{pmatrix},
\end{equation}
\end{widetext}
where material constants $l_1$, $l_2$ $m_1$, $m_2$, $n_1$, and $n_2$ are expressed via conventional  deformation potential tensor components as follows:~\cite{bir74,chuang96,winkelnkemper06}
\begin{equation}
\begin{split} \label{lmn_thru_D}
&l_1 = D_2 +D_4 +D_5, \quad m_1 = D_2 +D_4 -D_5,\\
&n_1 = 2 D_5,\quad l_2 = D_1, \quad m_2 = D_1+D_3,\\
&n_2 = \sqrt 2 D_6, \quad m_3 = D_2.
\end{split}
\end{equation}
We can also apply six fold rotational symmetry in the strain Hamiltonian of wurtzite structure which holds the relation  $l_1 - m_1 = n_1$.

The Hamiltonian  $\mathbf{\mathbf{H_{cv}}}$ of~(\ref{H-prim-1}) ($\mathbf{H^\dag _{cv}}$ is its Hermitian conjugate) can be written as
\begin{equation}
\mathbf{H_{cv}} = \left(H_{cx}~~  H_{cy}~~  H_{cz}\right),
\label{Hcv}
\end{equation}
where
\begin{equation}
H_{cx} = i P_2 k_x,~~ H_{cy} = iP_2 k_y,~~H_{cz} =\ iP_1 k_z.
\label{Hcx}
\end{equation}

\begin{figure*}
\includegraphics[width=17cm,height=7cm]{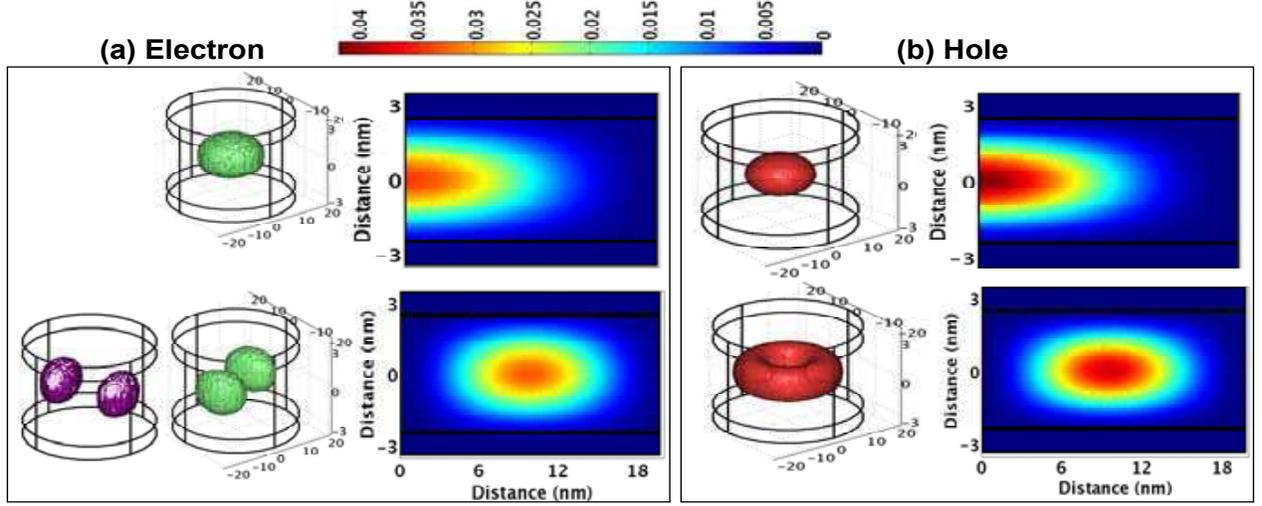}
\caption{\label{fig1}(Color online) Modeling of the distributions of electron and hole densities in AlN/GaN/AlN superlattice. Upper panel  shows the ground state wavefunctions and lower panel shows the first excited state wavefunctions. First and third columns show the results of 3-dimensional modeling in Cartesian coordinates while second and fourth columns show the results of 2-dimensional modeling in cylindrical coordinates. Notice that the first excited states of electrons in the conduction band are degenerate.  }
\end{figure*}
\begin{figure*}
\includegraphics[width=17cm,height=7cm]{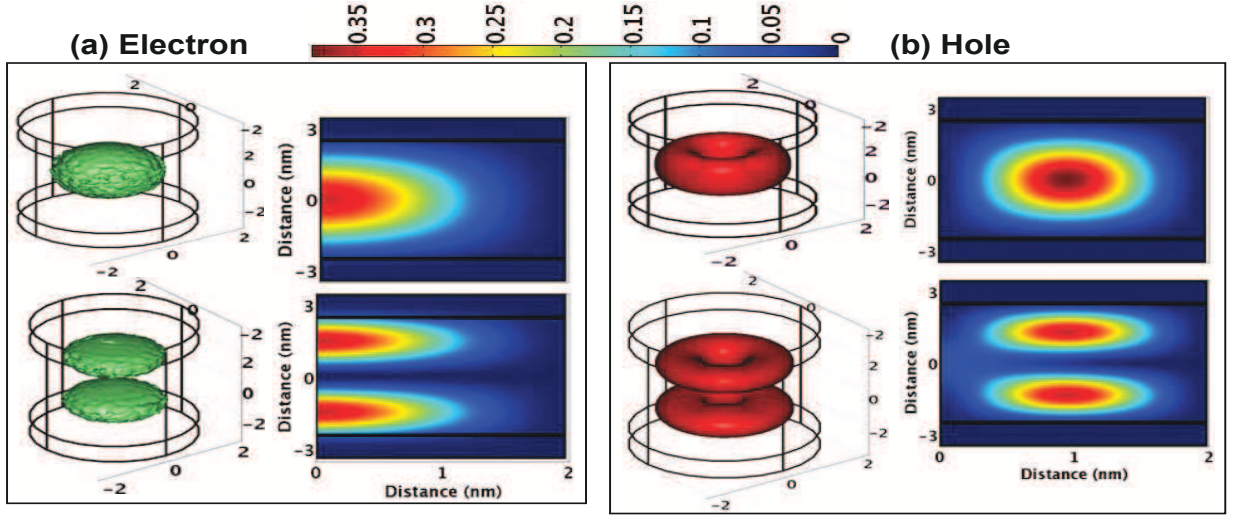}
\caption{\label{fig2}(Color online) Same as Fig.~\ref{fig1} but radius of the nanowire is chosen as 2 nm. By comparing Figs.~\ref{fig1} and \ref{fig2}, we clearly see the finite size effect on the localization of the distributions of electrons and holes wavefunctions.   }
\end{figure*}
\begin{figure*}
\includegraphics[width=18cm,height=8cm]{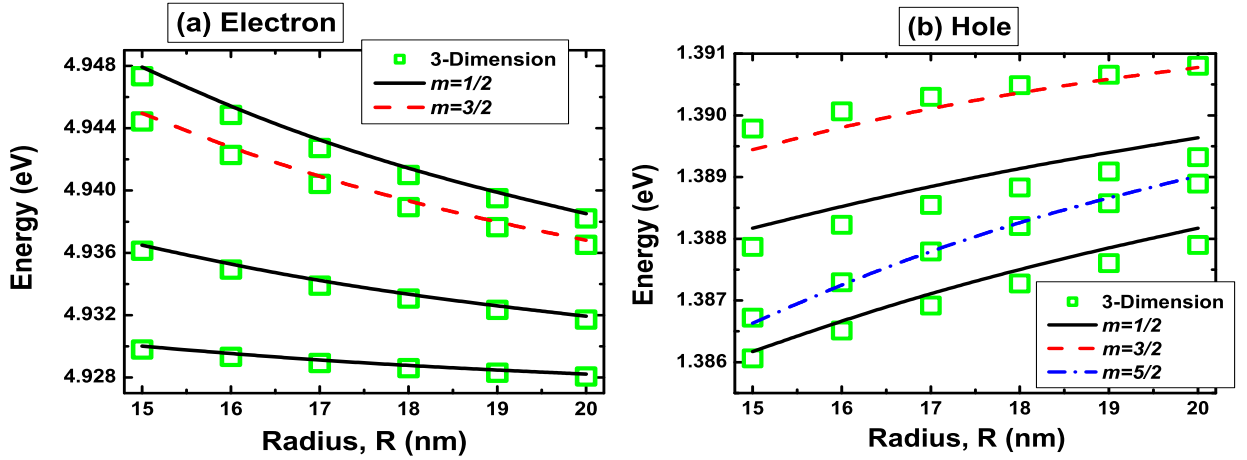}
\caption{\label{fig3}(Color online) Several energy eigenvalues of the lowest conduction and valence band states as a function of radius in a cylindrical AlN/GaN/AlN NWSLs. Eigenvalues obtained from the cylindrically symmetric 3-dimensional model in Cartesian coordinates  are in excellent agreement with the  2-dimensional model in cylindrical coordinates.}
\end{figure*}
\begin{figure*}
\includegraphics[width=12cm,height=8cm]{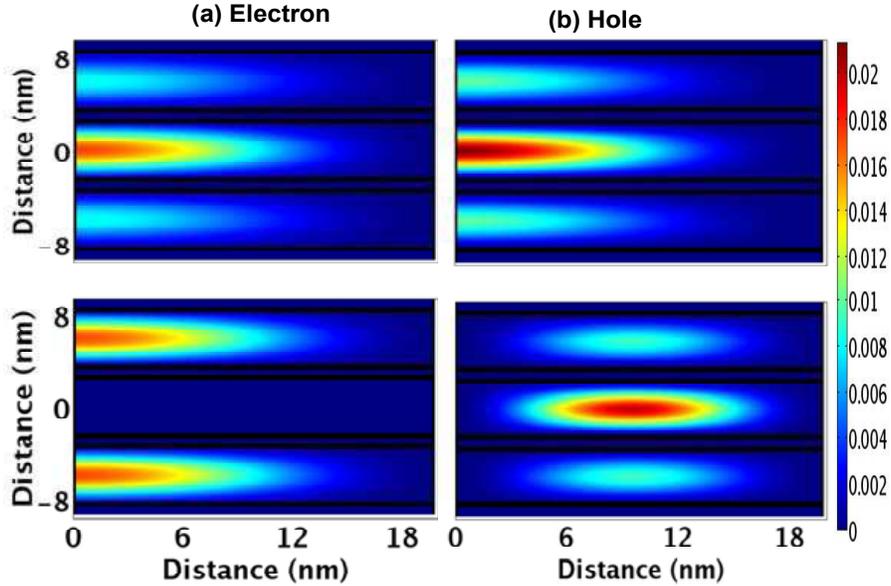}
\caption{\label{fig4}(Color online) Modeling of the distributions of electron and hole densities in three alternating layers of AlN/GaN superlattice. Upper panel  shows the ground state wavefunctions and lower panel shows the first excited state wavefunctions.}
\end{figure*}
\begin{figure*}
\includegraphics[width=18cm,height=10cm]{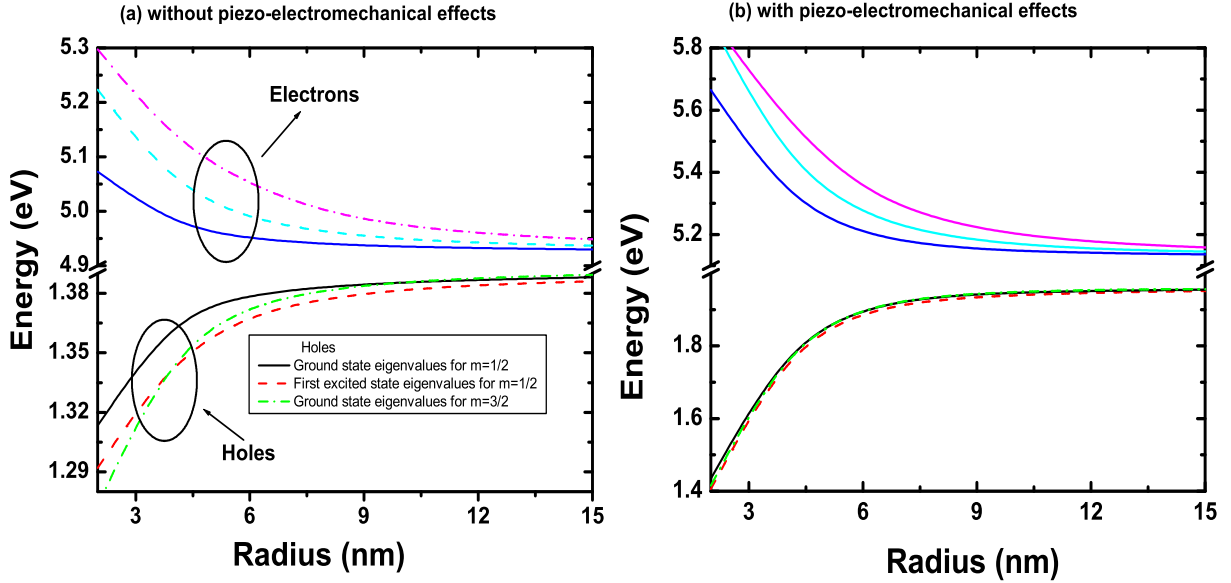}
\caption{\label{fig5}(Color online) Several energy eigenvalues of the lowest conduction and valence band states as a function of radius in  2-dimensional cylindrical AlN/GaN/AlN NWSLs.}
\end{figure*}
\begin{figure*}
\includegraphics[width=18cm,height=8cm]{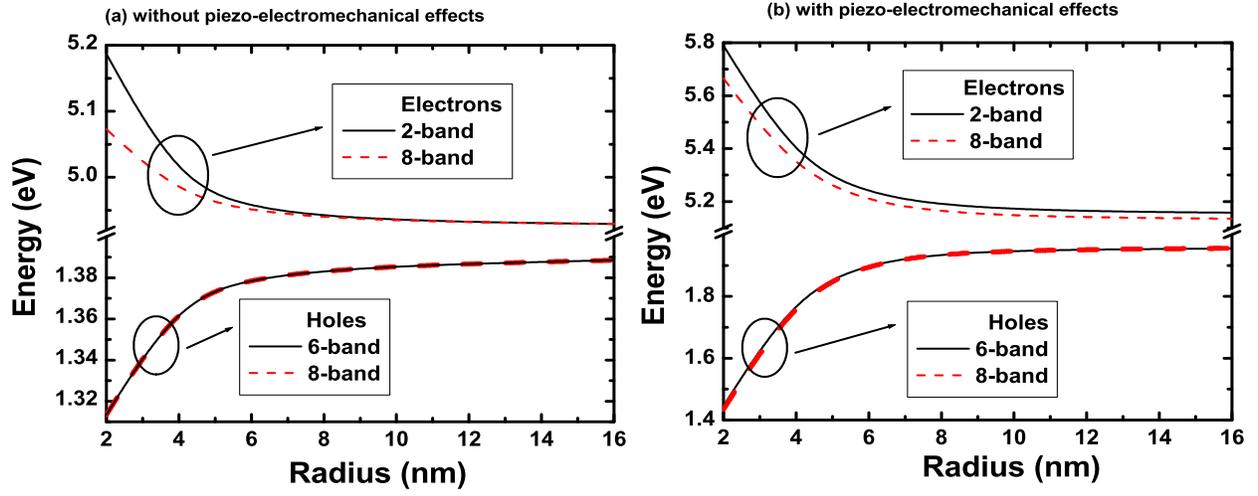}
\caption{\label{fig6a}(Color online) Ground state energy eigenvalues vs radius of  2-dimensional cylindrical AlN/GaN/AlN NWSLs. Here we compare the band structures of the NWSLs obtained from decoupled 2 conduction band and 6 valence band to that of 8-band $\mathrm{k\cdot p}$ Hamiltonian. Here we chose $m=1/2$. }
\end{figure*}
\begin{figure*}
\includegraphics[width=14cm,height=8cm]{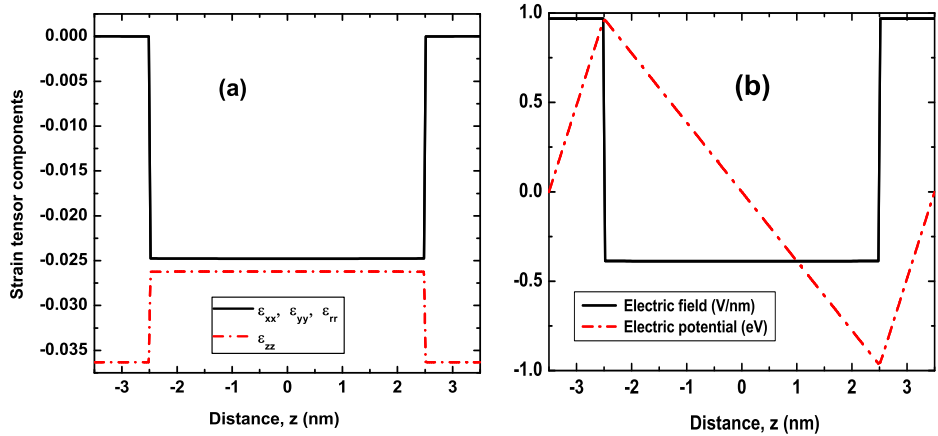}
\caption{\label{fig6}(Color online) (a) Strain tensor components for a AlN/GaN supperlattice. (b) Piezoelectric field and piezoelectric potential for a AlN/GaN supperlattice.  }
\end{figure*}
\begin{figure*}
\includegraphics[width=18cm,height=9cm]{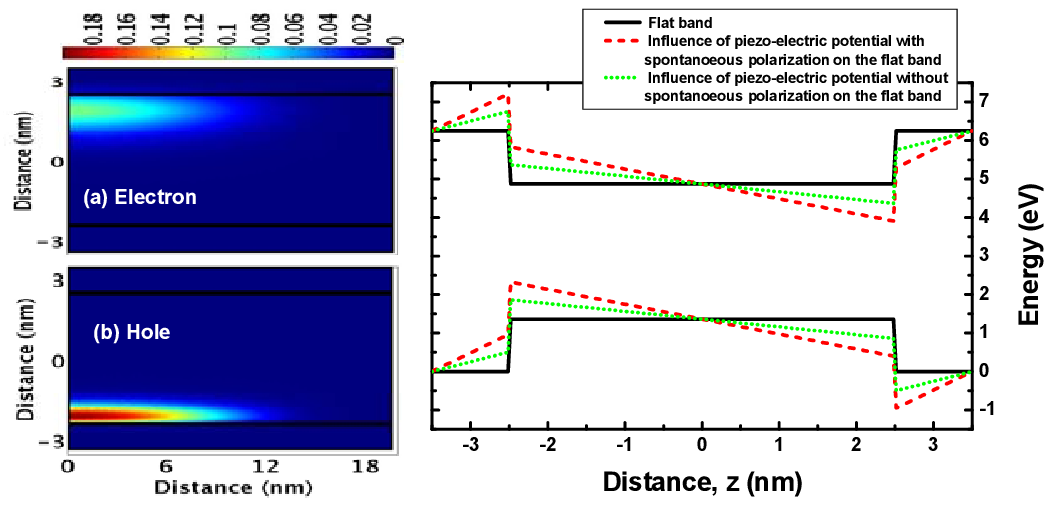}
\caption{\label{fig7}(Color online) Influence of piezo-electromechanical effects on the band structure calculations of wurtzite AlN/GaN supperllatice. }
\end{figure*}
\begin{figure*}
\includegraphics[width=18cm,height=9cm]{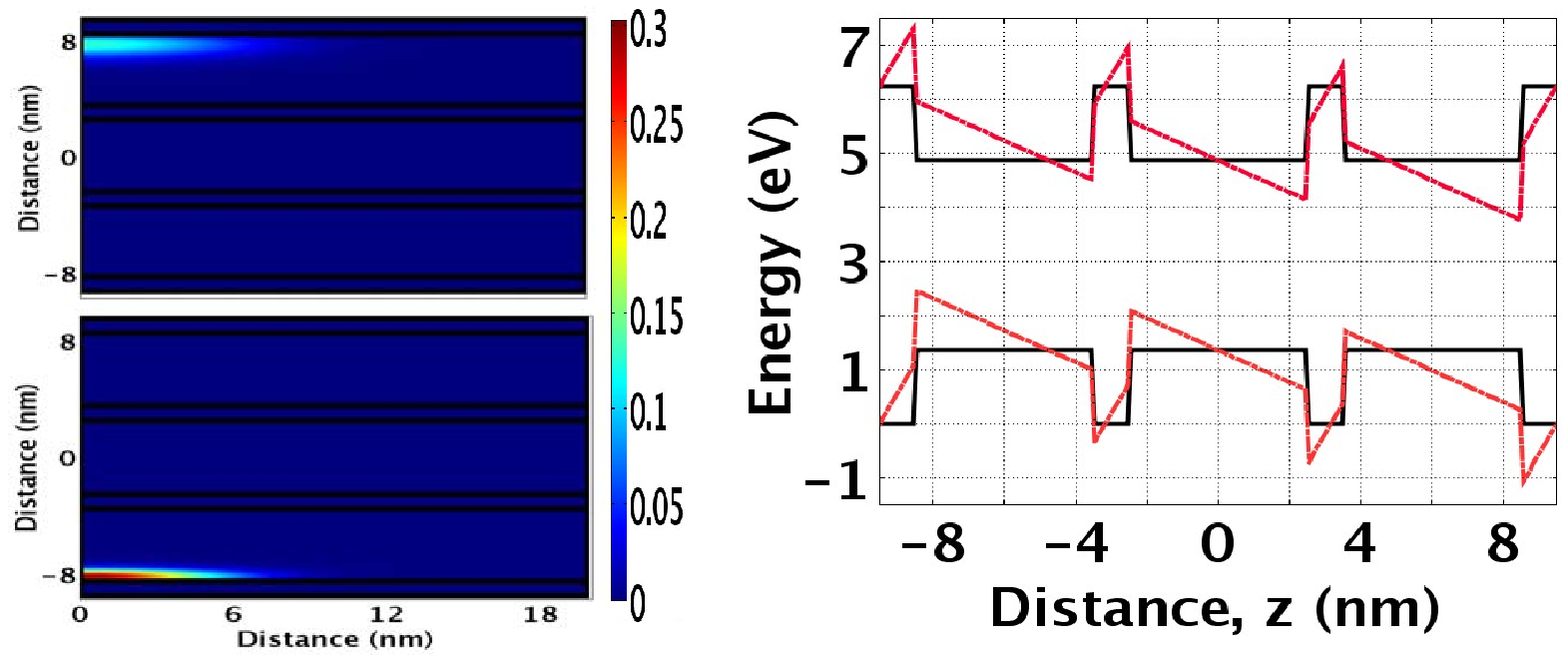}
\caption{\label{fig8}(Color online) Influence of piezo-electromechanical effects on the band structure calculations of wurtzite AlN/GaN supperllatice. }
\end{figure*}
\begin{figure*}
\includegraphics[width=16cm,height=8cm]{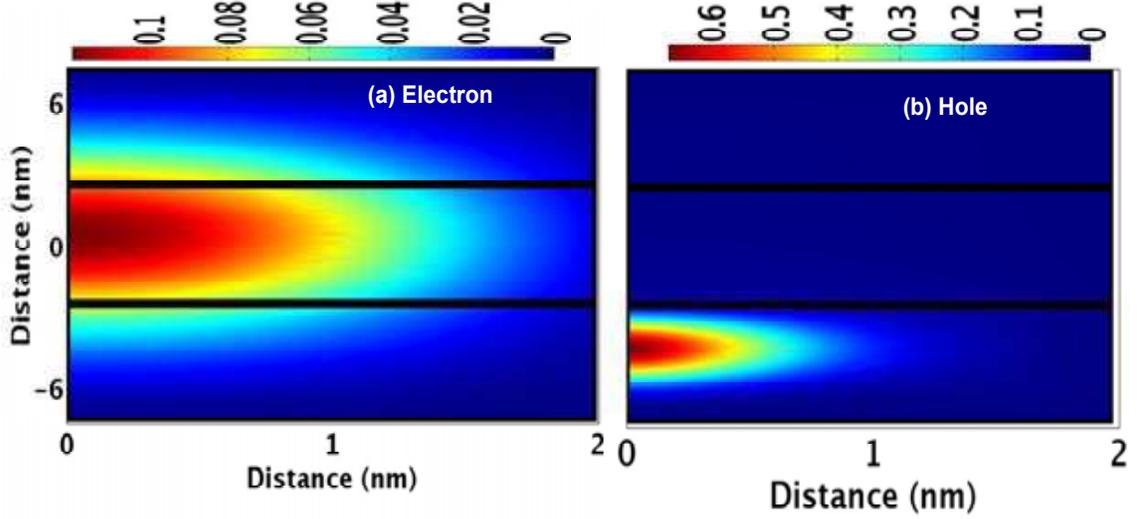}
\caption{\label{fig9}(Color online) Conduction band ground state (Fig.a) and valence band ground state (Fig.b) in symmetric $\mathrm{Al_xGa_{1-x}N}$/GaN/$\mathrm{Al_xGa_{1-x}N}$ NWSL structures with Al mole fraction x=0.01. Notice that the wavefunctions spread into the barrier material which indicates that the barrier material acts as an inversion layer.}
\end{figure*}
\begin{figure*}
\includegraphics[width=10cm,height=6cm]{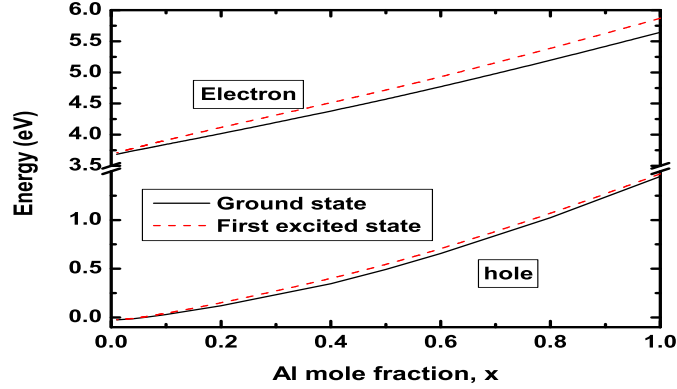}
\caption{\label{fig10}(Color online) Subband energy levels of electrons and holes vs  Al mole fraction  in symmetric $\mathrm{Al_xGa_{1-x}N}$/GaN/$\mathrm{Al_xGa_{1-x}N}$ NWSL structures.}
\end{figure*}
\begin{figure*}
\includegraphics[width=13cm,height=8cm]{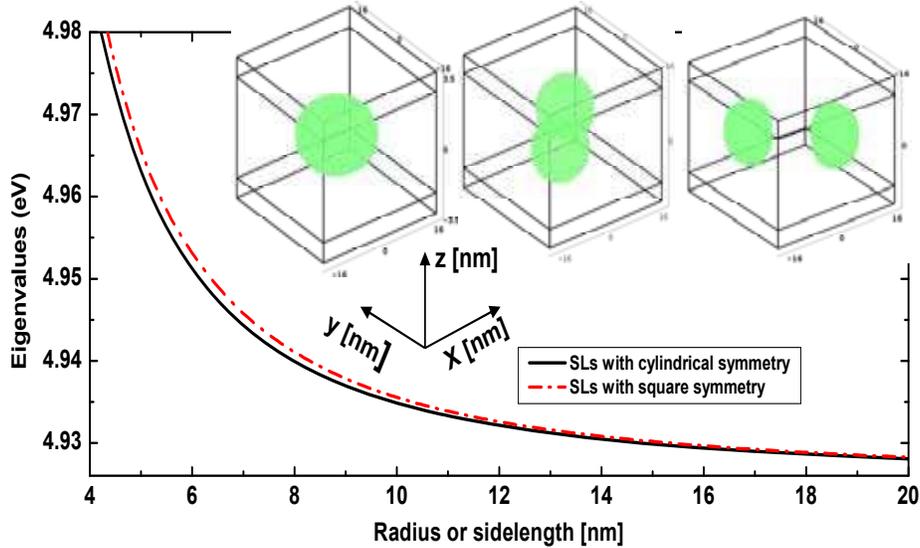}
\caption{\label{fig11}(Color online) Ground state eigenvalues vs radius of cylindrical NWSLs or sidelength of square NWSLs. Notice that the localization of first excited state weavefunction in square symmetry (inset plot) is different than for the case in cylindrical symmetry (\ref{fig1}(a) lower panel). }
\end{figure*}

\subsection{8-band $\mathbf{k\cdot p}$ model in cylindrical coordinates}
To derive the strain dependent 8-band $\mathbf{k\cdot p}$ model in cylindrical coordinates $\left(r, z\right)$, with $x = r \cos \phi$ and $y = r \sin \phi$, we introduce two different  unitary matrices as~\cite{prabhakar12a,takhtamirov13}
\begin{equation}
\label{S-1}
\mathbf{S_1} =
\begin{pmatrix}
1 & 0 & 0 & 0\cr
0 & \cos \phi & \sin \phi & 0\cr
0& -\sin \phi & \cos \phi & 0\cr
0& 0& 0 & 1
\end{pmatrix}.
\end{equation}
\begin{equation}
\label{S-2}
\mathbf{S_2} =
\begin{pmatrix}
e^{i\frac{\phi}{2}\sigma_z} & 0 & 0 & 0\cr
0 & e^{i\frac{\phi}{2}\sigma_z} & 0 & 0\cr
0& 0 & e^{i\frac{\phi}{2}\sigma_z} & 0\cr
0& 0& 0 & e^{i\frac{\phi}{2}\sigma_z}
\end{pmatrix}.
\end{equation}
We rotate the strain dependent 8-band $\mathbf{k\cdot p}$ Hamiltonian~(\ref{H-prim}) by $\mathbf{\widetilde{\widetilde{H}}}=\mathbf{S_2S_1HS^{-1}_1S^{-1}_2}$. We follow here the ideas first published in Ref.~\onlinecite{prabhakar12a,takhtamirov13}. Thus,  the eigenvalue problem~(\ref{H-prim}) can be written as
\begin{equation}
\mathbf{\widetilde{\widetilde{H}}}\mbox{\boldmath$\widetilde{\widetilde{\psi}}$}=E\mbox{\boldmath$\widetilde{\widetilde{\psi}}$}.
\label{H-prim-2}
\end{equation}
In Eq.~\ref{H-prim-2}, we  introduce new basis functions $\mathbf{\widetilde{f}}=\left(f_c ~f_r ~f_\phi ~f_z\right)$, where
\begin{equation}
f_r=\frac{xf_x+yf_y}{\sqrt{x^2+y^2}},~~~f_\phi=\frac{xf_y-yf_x}{\sqrt{x^2+y^2}}.
\end{equation}
Note that the new basis functions are invariant under the rotation of the Cartesian coordinate system around the z-axis i.e., $\mathrm{f}=\tilde{\mathrm{f}}S_1$. It can be seen that $f_r$ and $f_\phi$ do not obey the periodicity properties of the functions $f_x$ and $f_y$ and thus they are not the Bloch functions at this moment.
The total wavefunction $\mbox{\boldmath$\Psi$}$   and the envelope functions $\widetilde{\widetilde{\mbox{$\psi$} }}$ are related to each other through conventional basis functions $\mathbf{f}$ or the modified functions $\widetilde{ \mathbf{f}}$ ($\mathbf{f}=\widetilde{\mathbf{f}}\mathbf{S_1}$) as follows:
\begin{equation} \label{psi1}
\mbox{\boldmath$\Psi$} =  {\mathbf f} \mathbf S^{-1}_1 \mathbf S^{-1}_2  \widetilde{\widetilde{ \mbox{\boldmath $\psi$}}}={\mathbf{ \widetilde{f}}} \mathbf {S^{-1}_2}  \widetilde{\widetilde{ \mbox{\boldmath $\psi$}}}.
\end{equation}
Earlier we have noticed that $f_r$ and $f_\phi$  are not periodic and they are not Bloch functions so  it is convenient to retain the old basis functions $\mathrm{f}$ which are periodic and orthonormal. Thus, we obtain the old envelope functions $\psi$ as
\begin{equation} \label{psi2}
\psi = \mathbf{ S^{-1}_1} \mathbf{ S^{-1}_2}  \widetilde{\widetilde{ \mbox{\boldmath $\psi$}}}.
\end{equation}

To transform the Hamiltonian~(\ref{H-prim-1}) ($\mathbf{\widetilde{\widetilde{H}}}=\mathbf{S_2S_1HS^{-1}_1S^{-1}_2}$), we use the following identities:
\begin{equation}
\begin{split}
& \partial_x = \cos \phi \, \partial _r
- \frac {\sin \phi } r \, \partial _\phi,\\
&\partial_y = \sin \phi \, \partial _r
+ \frac {\cos \phi } r \, \partial_ \phi,
\end{split}
\end{equation}
as well as the relations
\begin{equation}
\begin{split}
&\varepsilon_{xx}= \varepsilon_{rr} \cos^2 \phi + \varepsilon_{\phi\phi}\sin^2 \phi - \varepsilon_{r \phi} \sin 2\phi,\cr
&\varepsilon_{yy}= \varepsilon_{rr} \sin^2 \phi + \varepsilon_{\phi\phi}\cos^2 \phi + \varepsilon_{r \phi} \sin 2\phi,\cr
&\varepsilon_{xy}= \frac {\varepsilon_{rr} - \varepsilon_{\phi\phi} } 2 \sin 2 \phi + \varepsilon_{r \phi} \cos 2 \phi,\cr
&\varepsilon_{xz}= \varepsilon_{r z} \cos \phi - \varepsilon_{\phi z} \sin \phi ,\cr
&\varepsilon_{yz}= \varepsilon_{r z} \sin \phi + \varepsilon_{\phi z} \cos \phi.
\end{split}
\end{equation}
We obtain the elements of $\mathbf{\widetilde{H}}=\mathbf{S_1HS^{-1}_1}$ (see~\ref{Hprimek}) as follows:
\begin{eqnarray}
\widetilde{H}_c&=& -A'_1\partial^2_z-A'_2\left(\partial^2_r-\frac{1}{r^2}\mbox{\boldmath ${\partial}^2_{\phi}$}-\frac{1}{r}
\partial_{r}\right)+U_c\nonumber\\
&&+a_1\varepsilon_{zz}+a_2\left(\varepsilon_{rr}+\varepsilon_{\phi\phi}\right),
\end{eqnarray}
\begin{eqnarray}
 \widetilde {H}^{\prime(k)}_{11} &= &\cos \phi \, H^{\prime(k)}_{11} \cos \phi + \cos \phi\, H^{\prime(k)}_{21}\sin \phi\nonumber\\
 &&+ \sin \phi\,H^{\prime(k)}_{12} \cos \phi + \sin \phi\,H^{\prime(k)}_{22}\sin \phi\nonumber\\
&=&-L_1 \left( \partial_r^2 + \partial_r \frac 1 r \right) - M_1 \frac 1 {r^2} \partial_\phi^2
-M_2 \partial_z^2,\\
 \widetilde {H}^{\prime(k)}_{12} &= &-\cos \phi \, H^{\prime(k)}_{11} \sin \phi + \cos \phi\, H^{\prime(k)}_{21}\cos \phi \nonumber\\
 &&- \sin \phi\,H^{\prime(k)}_{12} \sin \phi + \sin \phi\,H^{\prime(k)}_{22}\cos \phi\nonumber\\
&=&-L_1 \partial_r \frac 1 r \partial_ \phi + M_1 \left( \frac 1 r \partial_r \partial_\phi + \frac 1 {r^2}\partial_\phi \right).
\end{eqnarray}
The rest of the elements are obtained analogously. Thus the full matrix of the kinetic energy Hamiltonian~(\ref{Hprimek})  in the rotated frame $\mathbf{\widetilde{H}}=\mathbf{S_1HS^{-1}_1}$ can be written as:
{\scriptsize
\begin{widetext}
\begin{equation} \label{tildeHk}
 \widetilde {\mathbf H}^{(k)} = -
\begin{pmatrix}
L_1 \left[ \partial_r^2 + \partial_r \frac 1 r \right] + M_1 \frac {\partial_\phi^2} {r^2}
+ M_2 \partial_z^2 &
L_1 \partial_r \frac {\partial_ \phi} r - M_1 \left[ \frac {\partial_\phi} r \partial_r + \frac {\partial_\phi} {r^2} \right] &
N_2 \partial_r \partial_z \\
L_1 \left[ \frac {\partial_\phi} r \partial_r + \frac {\partial_\phi} {r^2} \right]-
M_1 \partial_r \frac {\partial_ \phi} r &
M_1 \left[ \partial_r^2 + \partial_r \frac 1 r \right] + L_1 \frac {\partial_\phi^2} {r^2}
+ M_2 \partial_z^2 &
N_2 \frac {\partial_\phi} r \partial_z \\
N_2 \left[ \partial_r +\frac 1 r \right]\partial_z &
N_2 \frac {\partial_\phi} r \partial_z &
M_3 \left[ \partial_r^2 + \frac 1 r \partial_r + \frac {\partial_\phi^2} {r^2} \right]+
L_2 \partial_z^2
\end{pmatrix}.
\end{equation}
\end{widetext}
}
Note  that $\partial _r ^{\rm T} = -\left( \partial_r + r^{-1}\right)$ and thus the Hamiltonian (\ref{tildeHk}) is Hermitian. It can be seen that Hamiltonian (\ref{tildeHk}) resembles the one in Refs.~\onlinecite{voon05} with different coefficients due to the fact that our choice of the basis functions and unitary rotation matrix are different.~\cite{takhtamirov13} Since $\widetilde {\mathbf H}^{(so)}=\mathbf{S_1H_{so}S^{-1}_1}$ (see~\ref{tildeHso}) will contain $\phi$ that does not commute with the operator $-i\partial_\phi$, we need to perform another set of rotation $ \widetilde {\widetilde {\mathbf H}}\,\!^{(so)}=\mathbf{S_2\widetilde{H}S^{-1}_2}$ to remove the $\phi$ dependency  from  $\widetilde {\mathbf H}^{(so)}$. Thus it is also required to find all the elements of $\mathbf{\widetilde{H}}=\mathbf{S_1HS^{-1}_1}$ under the rotation $\mathbf{\widetilde{\widetilde{H}}}=\mathbf{S_2\widetilde{H}S^{-1}_2}$.
Again, by performing another lengthy algebraic transformations, we obtain the elements of $\mathbf{\widetilde{\widetilde{H}}}=\mathbf{S_2\widetilde{H}S^{-1}_2}$ (see~(\ref{H-prim-1})). For example, ${\widetilde{\widetilde{H}}_c}$ of Eq.~(\ref{Hc}) can be written as
\begin{eqnarray}
{\widetilde{\widetilde{H}}_c}&=& -A'_1\partial^2_z-A'_2\left(\partial^2_r-\frac{1}{r^2}\mbox{\boldmath $\hat{\partial}^2_{\phi}$}-\frac{1}{r}
\partial_{r}\right)+U_c\nonumber\\
&&+a_1\varepsilon_{zz}+a_2\left(\varepsilon_{rr}+\varepsilon_{\phi\phi}\right),
\end{eqnarray}
where we use the identity:
\begin{eqnarray}
\mbox{\boldmath $\hat{\partial}_{\phi}$}&=&\mathrm e ^{ i \frac \phi 2 \sigma_z } \partial _\phi
\mathrm e ^{ -i \frac \phi 2 \sigma_z } = \partial _\phi - \frac i 2 \sigma_z\nonumber\\
&&=\begin{pmatrix}
\partial_\phi - \frac i 2&0\cr
0&\partial_\phi + \frac i 2
\end{pmatrix}.
\end{eqnarray}
Similarly, ${\widetilde{\widetilde{H}}_{cx}}$ of Eq.~(\ref{Hcx}) can be written as
\begin{equation}
{\widetilde{\widetilde{H}}_{cx}}=P_2\left(\partial_r+\frac{1}{r}\right).
\end{equation}
Also, first element of $\widetilde {\widetilde {\mathbf H}}\,\!^{k}=\mathbf{S_2S_1\mathbf{H}^kS^{-1}_1S^{-1}_2}$ in~(\ref{Hprimek}) can be written as
\begin{equation}
\widetilde {\widetilde {\mathbf H}}\,\!^{k}_{11} =-L_1 \left( \partial_r^2 + \partial_r \frac 1 r \right) - M_1 \frac 1 {r^2} \mbox{\boldmath $\hat{\partial}^2_{\phi}$}-M_2 \partial_z^2.
\end{equation}
In a similar fashion, one can find the rest of the elements of strain dependent 8-band $\mathbf{k\cdot p}$ Hamiltonian in cylindrical polar coordinates ($r,\phi,z$). The results of this procedure were first reported in Ref.~\onlinecite{prabhakar12a}.

Following it, we can transform the  spin-orbit interaction Hamiltonian~(\ref{Hso}) as  $\widetilde {\widetilde {\mathbf H}}\,\!^{(so)} = \mathbf S_2 \widetilde{\mathbf H}^{(so)} \mathbf S_2 ^{-1}$, where $ \widetilde{\mathbf H}^{(so)} = \mathbf S_1 \mathbf H^{(so)} \mathbf S_1 ^{-1}$. Thus we can write
\begin{equation}
\widetilde {\mathbf H}^{(so)}= i
\begin{pmatrix}
0&-\Delta_2 \sigma_z&\Delta_3 \sigma_\phi\cr
\Delta_2 \sigma_z&0&-\Delta_3 \sigma_r\cr
-\Delta_3 \sigma_\phi&\Delta_3 \sigma_r&0
\end{pmatrix},\label{tildeHso}
\end{equation}
where $\sigma_r = \sigma_x \exp \left( i \phi \sigma_z \right)$ and $\sigma_\phi = \sigma_y \exp \left( i \phi \sigma_z \right)$.

Notice that the spin-orbit Hamiltonian~(\ref{tildeHso}) depends on $\phi$ which does not commute with the operator $-i\partial_\phi$. To avoid this dependence, we note  that
\begin{equation}
\mathrm e ^{ i \frac \phi 2 \sigma_z }\sigma_r
\mathrm e ^{ -i \frac \phi 2 \sigma_z } = \sigma_x,\quad
\mathrm e ^{ i \frac \phi 2 \sigma_z }\sigma_\phi
\mathrm e ^{ -i \frac \phi 2 \sigma_z } = \sigma_y,
\end{equation}
and then the spin-orbit interaction Hamiltonian $\widetilde {\widetilde {\mathbf H}}\,\!^{(so)}$ transforms into  its initial form~(\ref{Hso}):
\begin{equation} \label{tildetildeHso}
\widetilde {\widetilde {\mathbf H}}\,\!^{(so)} = \mathbf{S_2} \widetilde {\mathbf H}^{(so)} \mathbf{S^{-1}_2} = {\mathbf H}^{(so)}.
\end{equation}
For cylindrically symmetric  systems, the total rotated Hamiltonian $\widetilde {\widetilde {\mathbf H}}$ commutes with the z-component of the total  angular momentum operator $j_z=-i\hbar\partial_\phi$. These two commuting operators have the common eigenfunctions so that the total rotated envelope functions can be chosen in the form of
\begin{equation} \label{tildetildeF}
\widetilde {\widetilde {\mbox{\boldmath $\psi$}}} = \frac {{\mathrm e}^{im\phi}}{\sqrt {2\pi}}\, \mbox{\boldmath $\psi$} \left(r, z \right),
\end{equation}
where  $m = \pm 1/2, \pm 3/2, \dots$ are the eigenvalues of the z-components of the total angular momentum $j_z$. The advantages of using these basis functions~(\ref{tildetildeF}) in the rotated frame is that the total Hamiltonian becomes  $\phi$ independent from  the strain dependent 8-band $\mathbf{k\cdot p}$  Hamiltonian.

Now we summarize the strain dependent 8-band $\mathbf{k\cdot p}$  Hamiltonian in cylindrical coordinates as follows:
{\scriptsize
\begin{widetext}
\begin{equation} \label{tildeHkm}
\widetilde {\widetilde {\mathbf H}}=
\begin{pmatrix}
\begin{pmatrix}
-A'_2\left(\partial^2_r-\frac{\mathbf{m}^2}{r^2}-\frac{1}{r}
\partial_{r}\right)\\-A'_1\partial^2_z+U_c+a_1\varepsilon_{zz}+\\a_2\left(\varepsilon_{rr}+\varepsilon_{\phi\phi}\right)
\end{pmatrix}\mathbf{\hat{1}}
&\mathbf{\hat{1}} P_2\left(\partial_r+\frac{1}{r}\right)&\frac{i \mathbf{\mathbf{m}} P_2}{r}&\mathbf{\hat{1}}P_1\partial_z\\
-\mathbf{\hat{1}}P_2\partial_r &
\begin{pmatrix}
-L'_1 \left[ \partial_r^2 + \partial_r \frac 1 r \right] +\\ M_1 \frac {{\mathbf m}^2} {r^2}
- M_2 \partial_z^2\\
+U_{v6}+l_1\varepsilon_{rr}+\\
m_1\varepsilon_{\phi\phi}+m_2\varepsilon_{zz}
\end{pmatrix}\mathbf{\hat{1}}
&
\begin{pmatrix}
 i M_1 \left[ \frac {{\mathbf m}} r \partial_r + \frac {{\mathbf m}} {r^2} \right] \\
-i L'_1 \partial_r \frac {{\mathbf m}} r-i\triangle_2\sigma_z
\end{pmatrix}
&
\begin{pmatrix}
-N'_2 \partial_r \partial_z+n_2\varepsilon_{rz}\\
+i\triangle_3\sigma_y
\end{pmatrix}\mathbf{\hat{1}}
\\
- \frac{i \mathbf{\mathbf{m}} P_2}{r}&
\begin{pmatrix}
-i L'_1 \left[ \frac {{\mathbf m}} r \partial_r + \frac {{\mathbf m}} {r^2} \right]+\\
i M_1 \partial_r \frac {{\mathbf m}} r +i\triangle_2\sigma_z
\end{pmatrix}
&
\begin{pmatrix}
-M_1 \left[ \partial_r^2 + \partial_r \frac 1 r \right] + \\
L'_1 \frac {{\mathbf m}^2} {r^2}- M_2 \partial_z^2\\
+U_{v6}+m_1\varepsilon_{rr}\\
+l_1\varepsilon_{\phi\phi}+m_2\varepsilon_{zz}
\end{pmatrix}\mathbf{\hat{1}}
&
-i N'_2 \frac {{\mathbf m}} r \partial_z-i\triangle_3\sigma_x \\
-\mathbf{\hat{1}} P_1\partial_z &
\begin{pmatrix}
-N'_2 \left[ \partial_r +\frac 1 r \right]\partial_z \\
+n_2\varepsilon_{rz}-i\triangle_3\sigma_y
\end{pmatrix}\mathbf{\hat{1}}
&
-i N'_2 \frac {{\mathbf m}} r \partial_z+i\triangle_3\sigma_x &
\begin{pmatrix}
-M_3 \left[ \partial_r^2 + \frac 1 r \partial_r - \frac {{\mathbf m}^2} {r^2} \right]\\
-L'_2 \partial_z^2+U_{v1}+l_2\varepsilon_{zz}\\
+m_3\left(\varepsilon_{rr}
+\varepsilon_{\phi\phi}\right)
\end{pmatrix}\mathbf{\hat{1}}
\end{pmatrix},
\end{equation}
\end{widetext}
}
where $\partial^T_r=-\left(\partial_r+r^{-1}\right)$ and the matrix ${\mathbf m}$ has the form:
\begin{equation}
{\mathbf m} =
\begin{pmatrix}
m - \frac 1 2 &0\cr
0& m +\frac 1 2
\end{pmatrix}.
\end{equation}
By identifying    $\varepsilon_{rr}+\varepsilon_{\phi\phi}=\partial_r u_r+u_r/r+2a$ (see Eqs.~\ref{varepsilon_rr} and ~\ref{varepsilon_rz}) in cylindrical coordinates, we verified that the total strain dependent 8-band $\mathbf{k\cdot p}$ Hamiltonian (\ref{tildeHkm}) in cylindrical coordinates are rotationally invariant with respect to the rotation around the c-axis.

\section{Computational Method:}
We have used the Finite Element Method (FEM)~\cite{comsol35} and solve the corresponding eigenvalue problem of fully strain dependent 8-band $\mathbf{k\cdot p}$  Hamiltonian in 3D Cartesian coordinates  and in 2D cylindrical coordinates. For 3D solutions (both electromechanical and band structure calculations), we have imposed Neumann  boundary conditions i.e., the continuity equation must hold at the heterojunction which is also referred to as the internal boundaries.  Dirichlet boundary conditions are imposed on the rest of the boundary. For cylindrically symmetric  2D model, we define the z axis to be perpendicular to the plane of the
quantum-well layer and  r axis  that lies in the quantum-well plane. For electromechanical parts, we have imposed $u_r=0$, $\partial_z u_z=0$ and $\partial_r V=0$ along the symmetry axis (i.e., at r=0) (for details, see Ref.~\onlinecite{barettin08}). Here, we also impose Neumann boundary conditions  at the internal boundaries and Dirichlet boundary conditions at the rest of the boundary. For the band structure calculations of 2D cylindrical nanowires, we have imposed Neumann boundary conditions along the symmetry axis as well as at the internal boundaries. We have imposed Dirichlet boundary conditions at the rest of the boundary by assuming that  the total Hamiltonian of the nanowire is rotationally invariant around z-direction. Finally, in 3D Cartesian coordinates and in  2D cylidnrical coordinates, we have used the corresponding normalization conditions
\begin{eqnarray}
\int_V |\Psi|^2 dx dy dz=1,\\
\int_s |\Psi|^2 dr dz=1.
\end{eqnarray}
The materials constants for our computation are taken from Refs.~\onlinecite{prabhakar12b,komirenko99,vurgaftman03} and listed in tables~\ref{WZ_LP} and~\ref{WZ_BS}.
\begin{table}[b]
\caption{\label{WZ_LP} Lattice parameters of wurtzite GaN and AlN used in computation. If not indicated differently, they are taken from Ref.~\onlinecite{vurgaftman03}. The dependence of the material constants for $\mathrm{Al_xGa_{1-x}N/GaN}$ on positions $\mathrm{x}$ are derived from the empirical expression according to the Ref.~\onlinecite{vurgaftman03}.
}
\begin{ruledtabular}
\begin{tabular}{lll}
Parameter & GaN & AlN\\
\colrule
$a_0$ (\AA) &  3.189 &  3.112\\
$c_0$ (\AA) &  5.185 &  4.982\\
$c_{11}$ (GPa)& 390 & 396\\
$c_{12}$ (GPa)& 145 & 137\\
$c_{13}$ (GPa)& 106 & 108\\
$c_{33}$ (GPa)& 398 & 373\\
$c_{44}$ (GPa)& 105 & 116\\
$P_{sp}$ (C/m$^2$)& -0.034 & -0.090\\
$e_{31}$ (C/m$^2$) & -0.30$^a$ & -0.25$^a$ \\
$e_{33}$ (C/m$^2$) & 1.06$^a$ & 1.79$^a$ \\
$e_{15}$ (C/m$^2$) & 0.33$^a$ & 0.42$^a$\\
$\kappa_1$ & 10.06$^b$ & 8.57$^b$\\
$\kappa_2$ &  9.28$^b$ & 8.67$^b$\\
\end{tabular}
\end{ruledtabular}
$^a$Ref.\onlinecite{prabhakar12b}. $^b$Ref. \onlinecite{komirenko99}.
\end{table}
\begin{table}[b]
\caption{\label{WZ_BS} Electron band structure parameters of wurtzite GaN and AlN used in computation. If not indicated differently, they are taken from Ref.~\onlinecite{rinke08}. The dependence of the material constants for $\mathrm{Al_xGa_{1-x}N/GaN}$ on positions $\mathrm{x}$ are derived from the empirical expression according to the Ref.~\onlinecite{vurgaftman03}. }
\begin{ruledtabular}
\begin{tabular}{lll}
Parameter & GaN & AlN\\
\colrule
$E_g$ (eV) & 3.51 & 6.25\\
$\Delta_{cr}$ (eV) & 0.034 & -0.295\\
$\Delta_{so}$ (eV) & 0.017$^a$ & 0.019$^a$\\
$m_\parallel$ &  0.19$m_0$ & \\
$m_\perp$ &  0.21$m_0$ & \\
$A_1$ & -5.947 &  \\
$A_2$ & -0.528 &  \\
$A_3$ & 5.414 &  \\
$A_4$ & -2.512 &  \\
$A_5$ & -2.510 &  \\
$A_6$ & -3.202 &  \\
$A_7$ (eV\AA) & 0.046 &  \\
$P_1$ (eV\AA) & 8.1 & \\
$P_2$ (eV\AA) & 7.9 & \\
$a_1$ (eV)  & -4.9$^a$   & -3.4$^a$\\
$a_2$ (eV)  & -11.3$^a$  & -11.8$^a$\\
$D_1$ (eV)  & -3.7$^a$   & -17.1$^a$\\
$D_2$ (eV)  &  4.5$^a$   &  7.9$^a$\\
$D_3$ (eV)  &  8.2$^a$   &  8.8$^a$\\
$D_4$ (eV)  & -4.1$^a$   & -3.9$^a$\\
$D_5$ (eV)  & -4.0$^a$   & -3.4$^a$\\
$D_6$ (eV)  & -5.5$^a$   & -3.4$^a$\\
\end{tabular}
\end{ruledtabular}
$^a$Ref. \onlinecite{vurgaftman03}.\\
\end{table}

\section{Results and Discussions}

We have plotted probability distributions of the ground and first excited states  of electrons and holes in Fig.~\ref{fig1}(a) and Fig.~\ref{fig1}(b), respectively. Here we consider the radius of the AlN/GaN/AlN nanowire as 20 nm. Also, in Fig.~\ref{fig2}(a) and Fig.~\ref{fig2}(b), we have plotted probability distributions of the ground and first excited states of electrons and holes where the radius  of the AlN/GaN/AlN NWSLs was only 2 nm in order to investigate the radial influence of the localization of the electron and hole wavefunctions.
By comparing the distribution functions of the ground state of electron in Fig.~\ref{fig1}(a) for R=20 nm and  in Fig.~\ref{fig2}(a) for R=2 nm, we  have found
the maximum probability point at (r,z)=(0,0). However, for the excited states, we have found the zero probability point at  $(r,z)=(0,0)$ and maximum probability points at $(r,z)=(0,\pm 1.5)$.
Also for hole states, we  have found the maximum probability   point at (r,z)=(0,0) for the ground state wavefunction in Fig.1(b) (R=20 nm) and  zero probability  point at  $(r,z)=(0,0)$ for the ground state wavefunction in Fig.2(b) (R=2 nm).
However, for hole  excited states, we have found the  zero probability point at  $(r,z)=(0,0)$ and two zero probability  points at $(r,z)=(0,\pm 1.5)$.
In Fig.~\ref{fig3}, we compare  several eigenvalues  of electron and hole states obtained from the 3D model in Cartesian coordinates  to those values obtained from the 2D model in cylindrical coordinates.
It can be seen that the eigenvalues obtained from the 3D and 2D models are in excellent agreement with difference less than $5\%$. The numerical error is due to the FEM implementation and can be reduced further by refining the mesh in 3D. This comparison of  eigenvalues in 3D and 2D models demonstrates that the derived 8-band $\mathbf{k\cdot p}$ model in cylindrical coordinates can be used for NWSLs instead of general 3D model.
This substantially reduces the required computational time.
Notice that for the holes in the valence band, the lowest state of $m=\pm 3/2$ corresponds to the ground state (see dashed line (red) in Fig.~\ref{fig3}(b)) and the lowest state of $m=\pm 1/2$ corresponds to the first excited state (see solid line (black) in Fig.~\ref{fig3}(b)).
In Ref.~\onlinecite{voon04a}, the Sercel-Vahala basis~\cite{sercel90} was used for the model reductions in the case of cylindrical coordinates. The methodology proposed in this paper is different. Based on Fig.~\ref{fig5}, we further analyze the differences between these two approaches for NWSLs with small radii. Comparisons between 3D and 2D models are reported for the first time here for NWSLs, but the developed 8-band  $\mathbf{k\cdot p}$ model in cylindrical coordinates is applicable also to cylindrical quantum dots and other low dimensional nanostructures with cylindrical geometry as long as the systems are invariant around c-axis.
In Fig.~\ref{fig4}, we have potted the probability distributions of ground and first excited states  of electrons and holes in three layers of AlN/GaN NWSLs. We again have found that the maximum probability  point for the ground state wavefunctions of electrons and holes are at (r,z)=(0,0).

In Fig.~\ref{fig5}, we investigate the influence of piezo-electromechanical effects on the band structure calculations of electrons and holes in  cylindrical AlN/GaN/AlN NWSLs (Figs.~\ref{fig5}(a) and (b) present the results  without and with pizo-electromechanical effects, respectively). We have substituted $m=\pm1/2$ in~(\ref{tildeHkm}) and found the ground, first, second and so on excited states  of electrons in the conduction bands. We confirm that  for electrons in the conduction band, first excited states of $m=\pm 1/2$ correspond to the ground state of $m=\pm 3/2$ and second excited states of $m=\pm 1/2$ correspond to the ground states of $m=\pm 5/2$ and so on. For holes in the valence band in Fig.~\ref{fig5}(a), we demonstrate the finite  radius influence at $R=10$ nm where we find  the crossing of the eigenstates between  the lowest states of m=1/2 and m=3/2. This indicates that for $R<10$ nm, the lowest state eigenvalue with m=1/2 corresponds to the  ground state and the lowest state eigenvalue with $m=3/2$ corresponds to the  first excited state. These results have not been previously reported.

In Fig.~\ref{fig6a}, we provide additional justification of utilizing 8-band $\mathbf{\mathrm{k\cdot p}}$ Hamiltonian in wide band gap AlN/GaN/AlN  NWSLs. We compare the ground state eigenvalues obtained from the 8-band $\mathrm{k\cdot p}$ Hamiltonian to those of decoupled 2-conduction  and 6-valence bands envelope function methods. We see that the band structures of holes obtained from 6-bands and 8-bands $\mathbf{\mathrm{k\cdot p}}$ Hamiltonian both provide correct estimations. However, for electrons in the conduction band, the influence of non-parabolicity term (i.e., when including realistic values of  $P_1$ and $P_2$) in the effective mass approximation in the 8-band $\mathbf{\mathrm{k\cdot p}}$ Hamiltonian induces a significant contribution to the band structure of wide band gap AlN/GaN/AlN  NWSLs. By substituting $P_1=P_2= 0$  in (\ref{tildeHkm}), one can find decoupled 2-conduction and 6-valence band Hamiltonians. In Fig.~\ref{fig6a} (b), we have included the piezo-electromechanical effect and shown that the energy difference between the ground state eigenvalues of 2-band and 8-band is enhanced. Further enhancement can be achieved if we bring the minima and maxima of the conduction and valence bands closer with the application of the gate controlled electric fields along z-direction.~\cite{prabhakar12b}

Bulk GaN and AlN have different lattice constants so when  (0001) GaN NWSLs structure is grown, a significant lattice mismatch occurs at the interface between the GaN wells and AlN barriers.~\cite{chuang96,barettin08,patil09} The strain induced polarization along z-direction results  in strong built in piezoelectric fields and increased potential along z-direction. We solve the  Navier equations (\ref{del-j}) for stress and Maxwel's equations (\ref{del-i}) for piezoelectric fields  in 3D on the one hand, and 2D coupled Eqs.~(\ref{sigma-rra}), (\ref{sigma-rrb})  and (\ref{sigma-rr}) in  cylindrical coordinates on the other hand,  to investigate the piezo-electromechanical effects in wurtzite AlN/GaN/AlN (single layer of GaN) QWs. We assumed pseudomorphic strain conditions and plotted the nonvanishing position dependent strain tensor comments ($\varepsilon_{xx}$, $\varepsilon_{yy}$, $\varepsilon_{zz}$, $\varepsilon_{rr}$) in Fig.~\ref{fig6}(a) and electric field and potential in Fig.~\ref{fig6}(b) as a function of position along the z-direction. In Fig.~\ref{fig7}, we investigate the influence of piezo-electromechanical effects on the band structure calculation of wurtzite AlN/GaN NWSLs. We see that the influence of piezo-electromechanical effect pushes  the minima of the  conduction band at the top of the well and also pushes the maxima of the valence band at the bottom of the well. As a result, we find the localization of the electron wavefunction at the top of the well and hole wavefunction at the bottom of the well.
In Fig.~\ref{fig8}, we investigate the influence of piezo-electromechanical effects on the band structure calculation of wurtzite AlN/GaN multi layers of NWSLs. Here again, we see that the  wavefunction of electrons in  the conduction band is localized at the top of the GaN QW in the upper layer of the NWSLs  and the hole wavefunction is localized at the bottom of the GaN QW in the lower layer of the NWSLs.

We now turn to another key result of the paper: critical radius and quantum confinement in wurtzite NWSL structures.

The critical radius ($R_c > R$) corresponds to those values of the radius of NWSLs for which  the localization of the wavefunction penetrates into the barrier materials. For instance, if $R>R_c$, the localization of the wavefunction resides within the QW  materials. In  AlN/GaN NWSLs, the AlN barrier material acts as an infinite potential wall where the penetration of the wavefunction into the barrier material is not possible.
However, if we reduce the Al mole fraction in $\mathrm{\mathrm{Al_xGa_{1-x}N/GaN}}$ NWSLs, we also reduce the band offsets i.e., the barrier height of electrons (holes) is reduced at the interface of the heterojunction. In this situation, the influence of piezo-electromechanical effects is minimized and the penetration of electron (hole) wavefunctions can be seen to the barrier materials. In a  simple one band model within the effective mass approximation, the authors in Ref.~\onlinecite{voon04} provided the mathematical condition for the critical radius as
\begin{equation}
R \leq \left[\frac{m_B-m_w}{m_Bm_w V_0}\right]^{1/2} j_{mn}=R_c,
\end{equation}
where $j_{m,n}$ is the $\mathrm{n^{th}}$ zero of the Bessel function, $V_0$  is the barrier height and $m_B$, $m_w$  correspond to the effective masses of electron (hole) in the barrier and in the well. For $\mathrm{\mathrm{Al_xGa_{1-x}N/GaN}}$ with $x=0.01$, we find $V_0=8.82~\mathrm{meV}$, $m_B=0.2109m_0$ and  $m_w=0.21m_0$ for electrons in the conduction band which gives $R_c=6.3~\mathrm{nm}$.
In Fig.~\ref{fig9}, we plotted the ground state wavefunction of electrons (Fig.~\ref{fig9}(a)) and holes (Fig.~\ref{fig9}(b))   in the conduction  and valence bands respectively for the radius $R=2~\mathrm{nm} < R_c$. Notice that the wavefunctions of electrons and holes  spread into the barrier materials ($\mathrm{\mathrm{Al_xGa_{1-x}N/GaN}}$ with $x=0.01$). In Fig.~\ref{fig10}, we plotted the energy level diagram of ground and first excited states  of electrons and holes as a function of Al mole fraction. As we increase the Al mole fraction in $\mathrm{\mathrm{Al_xGa_{1-x}N/GaN}}$ NWSLs, we also enhance the influence of piezo-electromechanical effects. As a result, the subband energy difference between ground and first excited states of electron (hole) states also increases.

Finally, in Fig.~\ref{fig11}, we compare the band structures of AlN/GaN/AlN NWSLs of cylindrical and square symmetry. We consider the volume of the square and cylindrical NWSLs to be the same (i.e., $a=\sqrt\pi R$, where $a$ is the side length of the square NWSLs and $R$ is the radius of the cylindrical NWSLs). For smaller values of $R$ or $a$, the localized states are formed very close to the edge and thus we find the larger eigenvalues for the NWSLs of square symmetry.
However, for larger values of $R$ or $a$, the localized states are formed far away from the edge in both types of cylindrical and square NWSLs and thus we find the eigenvalues are not substantially influenced by the choice of either square or cylindrical symmetry. In this case,  the localization of first excited state weavefunction with square symmetry (inset plot of Fig.~\ref{fig11}) is different from the case with cylindrical symmetry (Fig.~\ref{fig1} (a) lower panel) which might indicate that the symmetry is broken in square shape NWSLs.

\section{Conclusions}
By applying rotationally invariant basis functions with  appropriate unitary transformation, we have  formulated and applied the strain dependent 8-band $\mathbf{k\cdot p}$ Hamiltonian in cylindrical coordinates. This approach allows us  to find the band structures of low dimensional semiconductor nanostructures in a computationally efficient way. This includes  quantum dots, quantum wells or nanowire supperlattices as long as the systems are cylindrically symmetric along z-direction. We have provided the detailed analysis  of the eigenvalues of electron (hole) states of wurtzite NWSLs for both 3D (Cartesian coordinates) and 2D (cylindrical coordinates) models.  In a single and vertically stacked multiple layer of NWSLs, we have shown that the  piezo-electromechanical effects push the electron wavefunction to the top of quantum well, while they push  the hole wavefunction to the bottom of the well. We have shown that the influence of piezo-electromechanical effects in $\mathrm{\mathrm{Al_xGa_{1-x}N/GaN}}$ structures can be minimized by varying the Al mole fraction. In a situation, where the radius of the NWSL is very small compared to the critical radius, the localization of the electron (hole)  wavefunction spreads into the barrier material. In this case, the barrier materials act as an inversion layer. Finally, we have shown the edge of the square symmetry enhances the eigenvalues of  the localized states for smaller values of  the side length of of the square NWSLs. For large NWSLs,  we have shown that the eigenvalues of  the localized states of square and cylindrical NWSLs  is not substantially  influenced by the edge states but the localization of  weavefunction with square symmetry  is different than for the case with cylindrical symmetry  which might indicate that the symmetry is broken in square shape NWSLs.

The authors acknowledge Dr. Sunil Patil for his input on the initial version of this paper. This work was supported by Natural Sciences and Engineering Research Council (NSERC) of Canada and Canada Research Chair (CRC) programs.


%

\end{document}